\documentclass[12pt]{iopart}
\usepackage{graphicx}
\usepackage{hyperref}
\usepackage{multirow}

\def\degC{$^\circ$C}
\def\WmK{Wm$^{-1}$K$^{-1}$}

\begin{document}


\title[Thermal conductivity of diamond-loaded glues for ATLAS]{Thermal conductivity of diamond-loaded glues for the ATLAS particle physics detector}

\author{E.A.\ Ouellette$^1$ and A.\ Harris$^2$}

\address{$^1$ University of Victoria\\
\ \ Victoria, British Columbia, Canada}
\address{$^2$ C-Therm Technologies\\
\ \ Fredericton, New Brunswick, Canada}

\ead{eao@uvic.ca, aharris@ctherm.com}

\begin{abstract}
The ATLAS experiment is one of two large general-purpose particle detectors at the Large Hadron Collider (LHC) at the CERN laboratory in Geneva, Switzerland.  ATLAS has been collecting data from the collisions of protons since December 2009, in order to investigate the conditions that existed during the early Universe and the origins of mass, and other topics in fundamental particle physics.  The innermost layers of the ATLAS detector will be exposed to the most radiation over the first few years of operation at the LHC.  In particular, the layer closest to the beam pipe, the B-layer, will degrade over time due to the added radiation.  To compensate for its degradation, it will be replaced with an Insertable B-Layer (IBL) around 2016.  The design of and R\&D for the IBL is ongoing, as the hope is to use the most current technologies in the building of this new sub-detector layer.  One topic of interest is the use of more thermally conductive glues in the construction of the IBL, in order to facilitate in the dissipation of heat from the detector.  In this paper the measurement and use of highly thermally conductive glues, in particular those that are diamond-loaded, will be discussed.  The modified transient plane source technique for thermal conductivity is applied in characterizing the glues across a wide temperature range.
\end{abstract}

\maketitle


\section{Introduction}

The ATLAS~\cite{atlas:2008a} experiment is one of six particle detectors around the 27 km circumference Large Hadron Collider (LHC) at the CERN laboratory in Geneva Switzerland.  ATLAS is a general purpose particle detector, whose goal is to distinguish and measure the energy of a wide variety of different particles arising from the collisions of protons, accelerated using the LHC.  The ultimate goal of ATLAS and the LHC is to investigate the conditions that existed during the early Universe and to better understand the origin of mass, as well as other topics in fundamental particle physics.  ATLAS has been collecting data since December 2009 at center of mass energies of 900 GeV and 7 TeV.  The LHC is slowly increasing its luminosity and the number of protons per bunch, both of which increases the collision rate.

\begin{figure}[h]
	\centering
	\includegraphics[width=.7\textwidth]{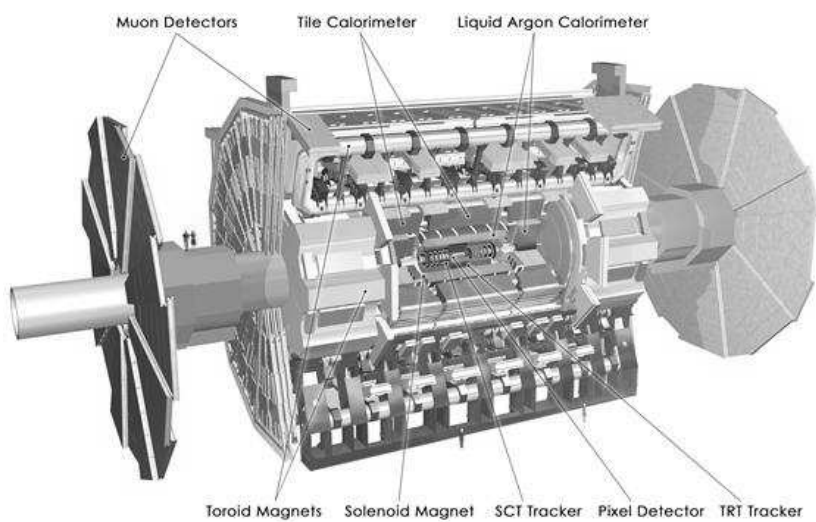}
	\caption{View of the ATLAS detector at CERN in Geneva Switzerland.}
	\label{fig:atlas}
\end{figure}

\begin{figure}[h]
	\centering
	\includegraphics[width=.7\textwidth]{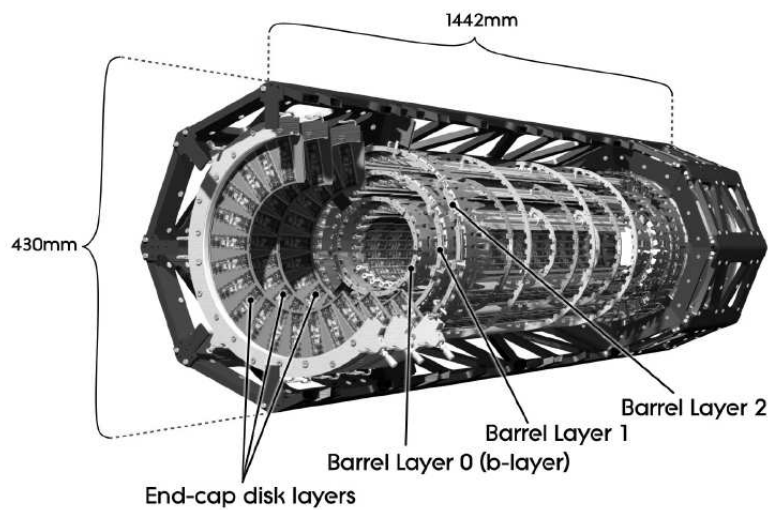}
	\caption{View of the ATLAS pixel system.}
	\label{fig:pixel}
\end{figure}

The ATLAS detector, pictured in Figure \ref{fig:atlas}, is comprised of different  subdetectors, each with a specific role: the Inner Detector for tracking and momentum measurements, the Calorimeters for energy determination and the Muon Spectrometers for muon identification.  The inner detector is also made up of three sub-systems: a transition radiation tracker, a semiconductor tracker and a pixel system.  The third, pictured in Figure \ref{fig:pixel}, is made up of three layers of pixel detectors and is the system closest to the LHC beam pipe.  The innermost of these three layers, the B-layer, will therefore be exposed to the most radiation over the first years of operation of the LHC and will reach its lifetime radiation dose of 500~kGy~\cite{atlas:2008b} before any other detector.  To compensate for its degradation over time, a replacement is planned for around 2016 for an insertable B-layer (IBL)~\cite{Flick:2010}.

\subsection{Insertable B-Layer}
The IBL will be integrated into the pixel system and is expected to last until the end of the LHC.  Much research and development are being put into designing the IBL to ensure that the most current technologies are utilized in its construction and it functions efficiently until the end of its mandate.

The basic layout of the IBL will be a layer of tilted pixel sensors, glued to a carbon foam base, with carbon fibre pipes passing through the foam for cooling purposes (using either liquid CO$_2$ or C$_3$F$_8$).  The system will run between \mbox{-40} and \mbox{-15~\degC}, thus it is essential that heat be effectively dissipated from the hot pixels and sensors.
One limiting component is the adhesive used to glue the pixels to the base, since the most commonly used epoxies have fairly low thermal conductivities (under 1~\WmK).
In simulations of the current design of the IBL, thermal conductivities of 1 to 2~\WmK\ are being used, since these were the values used in the original pixel detector~\footnote{Note that one of the main glues used in the original pixel detector was a thermal conducting, flexible epoxy glue, with a manufacturer quoted thermal conductivity of 1.97~\WmK~\cite{atlaspix:2009}.}.  Ideally, a value of 6~\WmK\ or higher would help in its design.
Fortunately, since the construction of the previous B-layer, more thermally conductive glues have become available.

Diamond has one of the largest thermal conductivities of any bulk material.  By loading glues with diamond powder, one can significantly increase its thermal conductivity.  For this paper, the thermal conductivity is measured for two AI Technology~\footnote{\url{www.aitechnology.com/}} glues, each made of the same epoxy base, though only one is diamond-loaded:
\begin{description}
	\item[$-$] ME7158, a stress-free, aluminum nitride-loaded glue, with a quoted thermal conductivity of 3.6~\WmK\ and
	\item[$-$] ME7159, a stress-free, diamond-loaded glue, with a quoted thermal conductivity of 11.4~\WmK.
\end{description}


\section{Experiment}

Thermal conductivity measurements were performed using the modified transient plane source technique employed by the C-Therm TCi Thermal Conductivity Analyzer.  The TCi system is comprised of an external sensor, control electronics and computer software. The sensor includes a central heater/sensor element in the shape of a spiral surrounded by a guard ring. The guard ring generates heat in addition to the spiral heater wire, approximating a one-dimensional heat flow from the perspective of the spiral platinum wire~\cite{Harris:2008}.  The difference between this configuration and the traditional hot-wire or transient plane source method is that the central heater and guard ring are supported on a backing material, which provides mechanical support, electrical insulation, and thermal insulation which enable a one-sided interfacial measurement and greatly enhance flexibility. The external sensor is rated from \mbox{-50} to \mbox{+200~\degC}\ and is sealed so that it can test solids, liquids, powders and pastes.  

\begin{figure}[h]
	\centering
	\includegraphics[width=.6\textwidth]{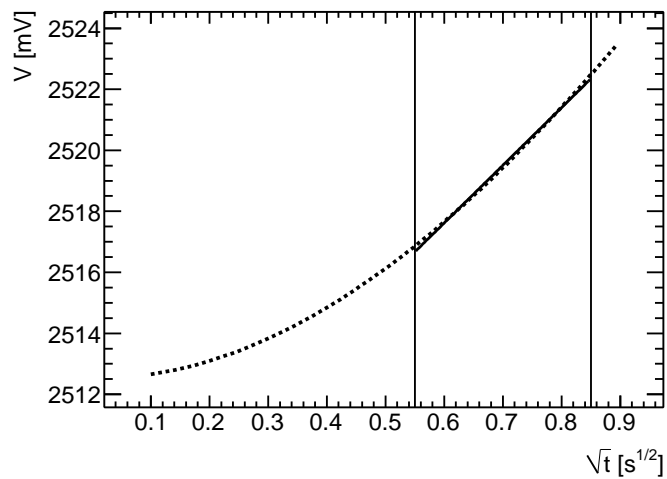}
	\caption{Representative voltage versus square root of time plot from a TCi thermal conductivity probe.}
	\label{fig:slope}
\end{figure}

The sample is tested by placing it in direct contact with the heating element of the sensor for a specific length of time (typically 0.8 seconds). A known current is applied to the sensorÕs heating element, providing a small amount of heat. This results in a rise in temperature at the interface between the sensor and the sample Ð typically less than \mbox{2~\degC}. This temperature rise at the interface induces a change in the voltage drop of the sensor's spiral heating element.  A typical voltage data chart is displayed in Figure \ref{fig:slope}~\cite{Harris:2005}. The slope of the voltage time chart is inversely proportional to the thermal conductivity of the sample material.  

The TCi is factory-calibrated with known standards across a broad thermal conductivity range from 0 (vacuum) up to 115~\WmK\ (yellow brass) for powders, liquids, foams, polymers, ceramics and metals.  
 
\begin{figure}[h]
	\centering
	\includegraphics[width=.5\textwidth]{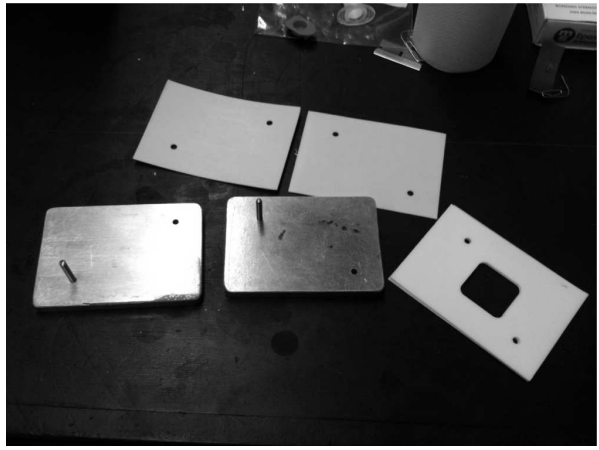}
	\caption{Machined teflon mold created to cure the glue into an appropriate shape for the measurement of its thermal conductivity.}
	\label{fig:mold}
\end{figure}

The minimum size requirements for the TCi are 2~mm thick and 18~mm diameter samples.  To meet these requirements, a 30~mm $\times$ 30~mm $\times$ 2~mm teflon mold was machined, pictured in Figure \ref{fig:mold}, to cure the glue into an appropriate shape.  Teflon was used to prevent the glue from sticking to its mold.  Thick pieces of aluminum were also used above and below the teflon mold in order to ensure pressure was applied evenly over the entire mold, creating a flat and homogeneous sample.

\section{Results}

\begin{figure}[h]
	\centering
	\includegraphics[width=.9\textwidth]{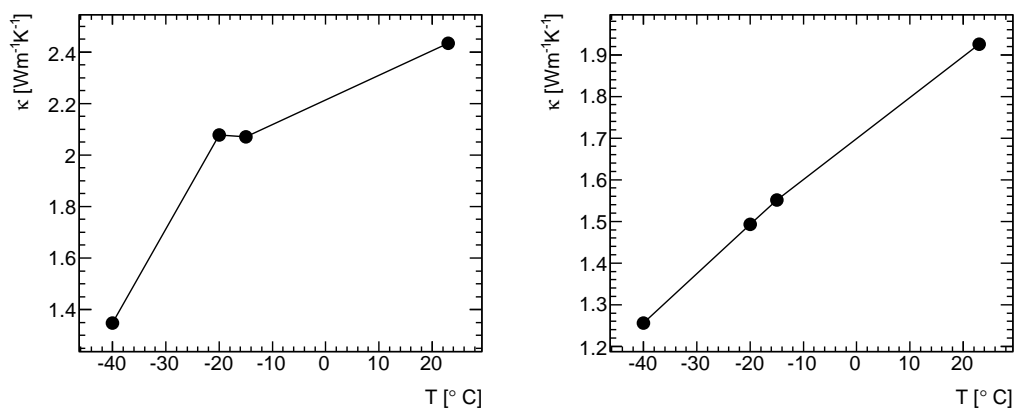}
	\caption{Plots showing the temperature dependence of the measurement of thermal conductivity ($\kappa$) of both glues: ME7158 on the left, ME7159 on the right.}
	\label{fig:plot}
\end{figure}

Since the IBL will run at temperatures in the range of \mbox{-40} to \mbox{-15~\degC}, the measurements were also done in this range.  One additional measurement was also done at ambient room temperature.  The results are tabulated in Table \ref{tab:results} and are plotted in Figure \ref{fig:plot}.  Note that the uncertainties are purely statistical.

\begin{table}[h]
	\caption{Results of the measurement of the thermal conductivity of both the ME7158 and ME7159 glues over the temperature range of \mbox{-40} to \mbox{+23~\degC}.}
	\label{tab:results}
	\begin{indented}
	\item[] \begin{tabular}{lcc}
		\br
		Glue	&Temperature [\degC]	&Thermal Conductivity [\WmK]\\
		\mr
		\multirow{4}{*}{ME7158}	&23	&2.433$\pm$0.003\\
		&-15	&2.070$\pm$0.004\\
		&-20	&2.078$\pm$0.002\\
		&-40	&1.347$\pm$0.001\\
		\mr
		\multirow{4}{*}{ME7159}	&23	&1.925$\pm$0.004\\
		&-15	&1.551$\pm$0.003\\
		&-20	&1.493$\pm$0.002\\
		&-40	&1.256$\pm$0.001\\
		\br
	\end{tabular}
	\end{indented}
\end{table}

\section{Analysis}

The measured thermal conductivities of both glues appear to be in disagreement with their manufacturer quoted values.  In particular, the diamond-loaded glue has a measured value less than its aluminum nitride-loaded partner.  After contacting AI Technology it was discovered that their measurements of thermal conductivity were performed at \mbox{121~\degC}\ in accordance with ASTM-C177.  Based on the plots in Figure \ref{fig:plot}, it does appear that the glues' thermal conductivity is proportional to temperature, though an extrapolation to \mbox{121~\degC}\ of a linear or quadratic fit would unlikely reach the quoted thermal conductivity for the ME7159 sample.  Unfortunately not enough data points are available to make a reasonable fit.

One possible source for this discrepancy could be in the curing process.  The glue itself is made of epoxy, resin and solvent.  During the curing process, the solvent will inevitably evaporate, leaving tiny air pockets within the glue.  Additional air pockets may also form simply by spreading the glue too quickly.  The combination of the glue and poorly thermally conductive air will lead to a mixture that has an effective thermal conductivity much less than that of the glue alone.

Possible contributors to the low conductivity of ME1759 are related to its diamond loading and the thickness of the bond line.  Upon magnification the concentration of diamond appears to be fairly low (less than 10\%) and may be insufficient to provide the higher thermal conductivity.  Although not discussed on the manufacturer's website, for a highly conductive material such as diamond in a much less conductive epoxy carrier, maximum effectiveness can be expected to occur for much thinner  bond lines where the diamonds act as a thermal bridge touching both sides.  Thus the sample thickness may have an effect on its thermal conductivity.

Another issue could be the method used  by the manufacturer to determine the glues' thermal conductivity.  The ASTM-C177 method is used primarily for insulators: ``The guarded-hot-plate apparatus is generally used to measure steady-state heat flux through materials having `low' thermal conductivities and commonly denoted as `thermal insulators'\ ''~\cite{astm:2004}.  There is ambiguity in the definition of `low' thermal conductivity, but these glues are clearly not insulators.  Thus it is possible that the technique used was not well suited to this material.

\section{Conclusion}

The hope for the IBL is to use glues which are highly thermally conductive in the range of \mbox{-40} to \mbox{-15~\degC}.  The two glues tested for this paper had measured values of $\kappa$ ranging from 1.35 to 2.07~\WmK\ (ME7158) and 1.26 to 1.55~\WmK\ (ME7159) over this temperature range.  This is much less than their quoted thermal conductivities at \mbox{121~\degC}.  Based on these results, use of either of these glues in the construction of the IBL should be carefully reconsidered in assessing the updated performance data with respect to the target thermal conductivity specifications for the material.

\section*{References}
\bibliographystyle{unsrt}
\bibliography{glue_thermal_cond}

\begin{thebibliography}{1}

\bibitem{atlas:2008a}
ATLAS Collaboration.
\newblock {The ATLAS Experiment at the CERN Large Hadron Collider}.
\newblock {\em JINST 3}, S08003, 2008.

\bibitem{atlas:2008b}
ATLAS Collaboration.
\newblock {ATLAS pixel detector electronics and sensors}.
\newblock {\em JINST 3}, P07007, 2008.

\bibitem{Flick:2010}
T.~Flick.
\newblock {IBL Ð ATLAS Pixel Upgrade}.
\newblock Technical Report ATL-INDET-PROC-2010-001. ATL-COM-INDET-2010-007,
  CERN, Geneva, Jan 2010.

\bibitem{atlaspix:2009}
ATLAS pixel collaboration.
\newblock {The ATLAS pixel detector mechanics and services}.
\newblock {\em submitted to JINST}, 2009.

\bibitem{Harris:2008}
A.~Harris.
\newblock {Measuring thermal conductivity}.
\newblock {\em Ceramic Industry - Special Report}, 2008.

\bibitem{Harris:2005}
A.~Harris and D.N. Sorensen.
\newblock {Thermal conductivity testing of minimal volumes of energetic
  powders}.
\newblock {\em Journal of Pyrotechnics}, 25, 2007.

\bibitem{astm:2004}
ASTM International.
\newblock {Test Method for Steady-State Heat Flux Measurements and Thermal
  Transmission Properties by Means of the Guarded-Hot-Plate Apparatus}.
\newblock C0177-04, 2004.

\end{thebibliography}

\end{document}